\documentclass[aps,prl,groupedaddress]{revtex4}

\usepackage{graphicx}
\begin{document}


\title{Optimized phase switching using a single
atom nonlinearity}

\author{Holger F. Hofmann $^{1,2}$}
\email{h.hofmann@osa.org}
\author{Kunihiro Kojima $^2$}
\author{Shigeki Takeuchi $^{1,2}$}
\author{Keiji Sasaki$^2$}
\affiliation{$^1$ PRESTO, 
Japan Science and Technology Corporation 
(JST)\\
$^2$ Research Institute for Electronic Science, Hokkaido 
University\\
Kita-12 Nishi-6, Kita-ku, Sapporo 060-0018, Japan}

\date{\today}

\begin{abstract}
We show that a nonlinear phase shift of $\pi$ can be obtained 
by using a single two level atom in a one sided cavity with 
negligible losses. 
This result implies that the use of a one sided cavity
can significantly improve the $\pi/18$ phase shift 
previously observed by Turchette et al. 
[Phys. Rev. Lett. {\bf 75}, 4710 (1995)].
\end{abstract}


\maketitle

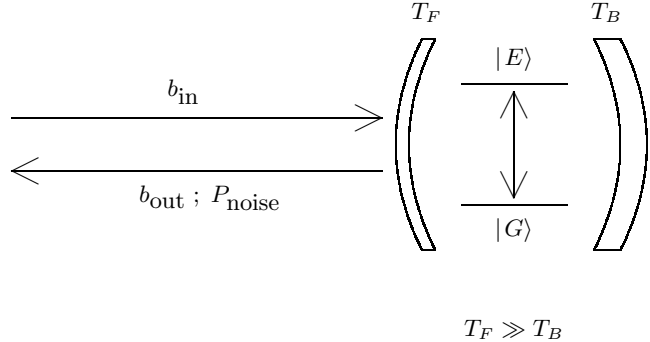
\begin{figure}
\begin{picture}(300,200)
\bezier{400}(240,60)(260,100)(240,140)
\bezier{400}(250,60)(270,100)(250,140)
\put(240,60){\line(1,0){10}}
\put(240,140){\line(1,0){10}}
\bezier{400}(180,60)(160,100)(180,140)
\bezier{400}(175,60)(155,100)(175,140)
\put(175,60){\line(1,0){5}}
\put(175,140){\line(1,0){5}}
\put(235,140){\makebox(20,20){$T_B$}}
\put(167,140){\makebox(20,20){$T_F$}}
\put(190,20){\makebox(40,20){$T_F \gg T_B$}}

\put(190,77){\line(1,0){40}}
\put(195,60){\makebox(30,15){$\mid \! G \rangle$}}
\put(190,123){\line(1,0){40}}
\put(195,125){\makebox(30,15){$\mid \! E \rangle$}}

\put(210,80){\line(0,1){40}}
\put(210,80){\line(1,2){5}}
\put(210,80){\line(-1,2){5}}
\put(210,120){\line(1,-2){5}}
\put(210,120){\line(-1,-2){5}}

\put(20,90){\line(1,0){140}}
\put(20,110){\line(1,0){140}}
\put(160,110){\line(-2,1){10}}
\put(160,110){\line(-2,-1){10}}
\put(20,90){\line(2,1){10}}
\put(20,90){\line(2,-1){10}}
\put(70,110){\makebox(30,20){$b_{\mbox{in}}$}}
\put(70,70){\makebox(50,20){$b_{\mbox{out}}\; ; \;P_{\mbox{noise}}$}}
\end{picture}
\caption{\label{geometry} Illustration of the dissipation 
free geometry for the realization of an optimized nonlinear 
phase shift at resonance. $T_F$ and $T_B$ denote the 
transmission coefficients of the two mirrors. Ideally, $T_B=0$.}
\end{figure}

One of the most significant achievements in the field of
quantum optics is the realization of strong nonlinear
effects by enhancing the coupling between single atoms
and the light field. In particular, the possibility
of obtaining large conditional phase shifts has attracted much
attention because of its potential usefulness in the 
realization of phase gates for optical quantum computation 
and similar manipulations of quantum states at the few 
photon level \cite{Tho98,Tur95,Nie,Wer99,Reb99,Poi92}. 
In order to optimize controlled phase shifts, it is 
desirable to avoid losses
while moving close to the resonance of the two level atom
causing the nonlinear phase shift. 
In the experiment by Turchette et al.\cite{Tur95}, 
the nonlinearity of the atom was observed in the phase
change of the light transmitted through the atom-cavity
system. However, the transmission of a two sided cavity
at the atomic resonance is very low, so the experiment 
was conducted at frequencies significantly detuned from 
this resonance. As a result, the phase shift observed
was limited to only about $\pi/18$.
In order to improve this phase shift, it is necessary
to move closer to resonance while avoiding dissipative losses.
In this paper, we therefore propose the use of a one sided
cavity with negligible losses to non-cavity modes.
In such a geometry, the total reflection is always close to
one and all dissipation is suppressed.  
The nonlinearity then has the maximal effect on the phase
of the light field while leaving the intensity unchanged.
This makes it possible to realize a nonlinear 
phase shift of $\pi$ at the atomic resonance.
Figure \ref{geometry} shows an illustration of this
dissipation free setup. Effectively, the cavity confines
the light field interacting with the two level atom to
a single beam with a well defined transversal profile.
The suppression of losses to non-cavity modes can be
achieved by covering a large solid angle of emission with
the confocal cavity mirrors. Further improvements may be 
possible by using dielectric materials, e.g. in a photonic 
crystal geometry.

The most simple case of this dissipation free setup is obtained in 
the bad cavity regime, where the cavity lifetime is so short that
the cavity field can be adiabatically eliminated. 
In terms of the conventional cavity quantum
electrodynamics parameters
this regime is characterized
by $\kappa \gg g$, where $\kappa$ is the cavity damping rate
and $g$ is the dipole coupling between the atom and the cavity. 
The effective dipole damping rate caused by emissions through 
the cavity is then given by $\Gamma = g^2/\kappa$ and the 
rate of spontaneous emission through the cavity is equal to
$2 \Gamma$. Note that this emission rate also includes any 
effects of cavity enhancement of spontaneous emission. 
If the transverse atomic decay rate to noncavity 
modes $\gamma$ is negligible ($\gamma \ll 2 g^2/\kappa$), 
nearly all emissions 
from the atom can be confined to the cavity and the total 
spontaneous emission rate of the excited atom in the cavity 
will be given by $2 \Gamma$.
With these parameters,
the relevant dynamics can be described by the well known
Bloch equations of the driven two level atom. For a two
level system with a ground state $\mid \! G \rangle$ and an
excited state $\mid \! E \rangle$, the Bloch equations can be
expressed in terms of the complex dipole operator
$\hat{\sigma}_-=\mid \! G \rangle \langle E \! \mid$ and
the inversion operator $\hat{\sigma}_z= 1/2 
(\mid \! E \rangle \langle E \! \mid - \mid \! G \rangle 
\langle G \! \mid)$. For a coherent driving field 
$b_{\mbox{in}}(t)$, the Bloch equations then read
\begin{eqnarray}
\label{eq:bloch}
\frac{d}{dt} \langle \hat{\sigma}_- \rangle
&=& -\Gamma \langle \hat{\sigma}_- \rangle + 
2 \sqrt{2 \Gamma}\; b_{\mbox{in}}(t)\; 
\langle \hat{\sigma}_z \rangle
\nonumber \\
\frac{d}{dt} \langle \hat{\sigma}_z \rangle
&=& - 2 \Gamma \left(\langle \hat{\sigma}_z \rangle 
                     + \frac{1}{2}\right) - 
\sqrt{2 \Gamma}\left(\; b^*_{\mbox{in}}(t)\; 
\langle \hat{\sigma}_- \rangle + \;b_{\mbox{in}}(t)\; 
\langle \hat{\sigma}_- \rangle^*\right).
\end{eqnarray}
Note that these equations of motion correctly describe the 
open system quantum dynamics of the two level atom. 
In particular, the effects of quantum fluctuations in the 
coherent driving field are fully included in the relaxation
dynamics of the atom. This can be seen clearly in the case
of $b_{\mbox{in}}=0$, where the incoming field is in the
vacuum state. The correct quantum mechanical description of
this situation is then given by spontaneous emission dynamics
that automatically include the dynamics of vacuum fluctuations
\cite{Wis93,Hof98,Hof98b}. Since the noise properties of a
coherent state are no different from the vacuum, the full 
quantum mechanical effect of a coherent driving field is 
indeed properly represented by equations (\ref{eq:bloch}).

The signal from the atom will now be observed in 
the reflected light, where it interferes with the directly
reflected component of the input field \cite{Foot}. 
The coherent
amplitude of the output field can then be described according 
to input-output theory \cite{Wal} as
\begin{equation}
\label{eq:inout}
b_{\mbox{out}}(t) = b_{\mbox{in}}(t) + \sqrt{2\Gamma} 
\langle \hat{\sigma}_- \rangle.
\end{equation}
The normalization of the light field amplitudes 
$b_{\mbox{in}}(t)$ and $b_{\mbox{out}}(t)$ has been chosen 
so that the squares of the amplitudes correspond to the 
photon current associated with the fields. 

Equations (\ref{eq:bloch}) and (\ref{eq:inout}) fully describe
the coherent dynamics of the field-atom system.
In particular, equation (\ref{eq:inout}) correctly describes 
the expectation value of the output field, as observed in
homodyne or heterodyne detection \cite{Wis93,Hof98,Hof98b}.
However, this is not sufficient to predict the average
emitted photon number, which can only be observed in precise
photon detection experiments \cite{Wis93}. This photon emission
rate also includes a noisy component $P_{\mbox{noise}}$ 
corresponding to random phase emissions. The total intensity 
of the emitted field can therefore be separated into a coherent
contribution $|b_{\mbox{out}}|^2$ and an incoherent 
contribution $P_{\mbox{noise}}$.
Energy conservation requires that any difference between
the output intensity and the input intensity
$|b_{\mbox{in}}|^2$ results in a corresponding change of the
atomic energy given by $\langle \sigma_z \rangle$. 
Therefore, the proper value of the emission noise in a
dissipation free geometry can be obtained from
\begin{equation}
\label{eq:energy}
\frac{d}{dt} \langle \hat{\sigma}_z \rangle
= |b_{\mbox{in}}|^2 - |b_{\mbox{out}}|^2 - P_{\mbox{noise}}.
\end{equation}
In order to identify the physical origin of noisy emissions,
it is useful to compare the total change in atomic energy
given by $-d/dt \langle \sigma_z \rangle$ with the change
in coherent field energy,
$|b_{\mbox{out}}|^2 - |b_{\mbox{in}}|^2$, 
\begin{equation}
\label{eq:power}
\begin{array}{ccccc}
-\frac{d}{dt} \langle \hat{\sigma}_z \rangle
&=& 2 \Gamma \left(\langle \hat{\sigma}_z \rangle 
                     + \frac{1}{2}\right) 
&+& 
\sqrt{2 \Gamma}\left(\; b^*_{\mbox{in}}(t)\; 
\langle \hat{\sigma}_- \rangle + \;b_{\mbox{in}}(t)\; 
\langle \hat{\sigma}_- \rangle^*\right)
\\
|b_{\mbox{out}}|^2 - |b_{\mbox{in}}|^2 
&=& 2 \Gamma |\langle \hat{\sigma}_- \rangle|^2 
&+& 
\sqrt{2 \Gamma}\left(\; b^*_{\mbox{in}}(t)\; 
\langle \hat{\sigma}_- \rangle + \;b_{\mbox{in}}(t)\; 
\langle \hat{\sigma}_- \rangle^*\right)
.
\end{array}
\end{equation}
The change in coherent field energy 
$|b_{\mbox{out}}|^2 - |b_{\mbox{in}}|^2$ depends only on 
the coherent dipole expectation value 
$\langle \hat{\sigma}_- \rangle$ and corresponds exactly to
the classical result for dipole emission without fluctuations. 
In the total change of energy at the atom, the square of 
the dipole expectation value $\langle \hat{\sigma}_- \rangle^2$
is replaced by $\langle \hat{\sigma}_z \rangle + 1/2$.
This replacement suggests that the difference $P_{\mbox{noise}}$ 
between the two terms can be interpreted as emission from dipole 
fluctuations, where $\langle \hat{\sigma}_z \rangle + 1/2$ 
represents the average of the squared dipole.
In terms of the atomic variables, the incoherent emission now
reads 
\begin{eqnarray}
\label{eq:noise}
P_{\mbox{noise}}
&=& \left(-\frac{d}{dt} \langle \hat{\sigma}_z \rangle \right) 
- \left(|b_{\mbox{out}}|^2 - |b_{\mbox{in}}|^2 \right)
\nonumber \\
&=& 2 \Gamma \left(\langle \hat{\sigma}_z \rangle 
    + \frac{1}{2} - |\langle \hat{\sigma}_- \rangle|^2
    \right).
\end{eqnarray} 
It is important to note that this result fundamentally limits
the coherence in the emission obtained from a single
two level atom, since the dipole expectation value is limited
by the maximal length of the Bloch vector, 
$|\langle \hat{\sigma}_- \rangle|^2 + 
\langle \hat{\sigma}_z \rangle^2 \leq 1/4$. 
With this limitation, the minimal incoherent emission is given by
\begin{equation}
\label{eq:limit}
P_{\mbox{noise}}\geq 2\Gamma \left(
\langle \hat{\sigma}_z \rangle + \frac{1}{2}\right)^2.
\end{equation}
Incoherent spontaneous emission is therefore an unavoidable
side effect of the saturation of a two level transition and
should be taken into account in the characterization of this
kind of nonlinearity.

It should be noted that the average rate of photon emission
can also be obtained using quantum trajectory theory 
\cite{Wis93}. However, the result of such an analysis is 
necessarily identical to the one obtained from the energy 
conservation argument applied above. Effectively, quantum 
trajectory theory does not modify the averaged results 
obtained in this paper. A more detailed analysis of the 
relationship between field and dipole fluctuations in the 
spontaneous emission of a two level atom mainly provides 
a time resolved description of the noise dynamics, showing 
that the minimal noise given by equation (\ref{eq:limit}) 
is related to unavoidable diffusion 
effects in the open system dynamics of the fully coherent atomic 
Bloch vector \cite{Hof98}. Quantum trajectory theory thus confirms 
the identification of $P_{\mbox{noise}}$ with time dependent 
dipole fluctuations in the atomic system suggested by 
equations (\ref{eq:noise}) and (\ref{eq:limit}). 

\begin{figure}
\begin{picture}(300,200)
\put(30,30){\makebox(250,150){\includegraphics{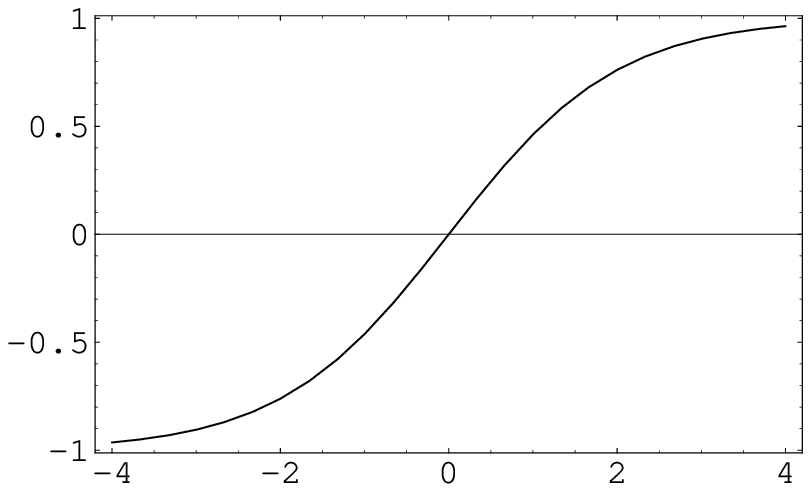}}}
\put(20,90){\makebox(20,40){\large 
$\frac{b_{\mbox{\small out}}}{b_{\mbox{\small in}}}$}}

\put(140,10){\makebox(60,20){\large $\log_{10} 
\left[\; 4 |b_{\mbox{\small in}}|^2 / \Gamma \;\right] $}}
\end{picture}
\caption{\label{response} Nonlinear response
of the dissipation free two level system given by the 
dependence of the amplitude ratio $b_{\mbox{out}}/b_{\mbox{in}}$ 
on the scaled input intensity 
$4 |b_{\mbox{in}}|^2 / \Gamma$.}
\end{figure}

The nonlinear response of the atomic system at resonance can
now be derived using equations (\ref{eq:bloch}) and 
(\ref{eq:inout}). If the input field is given
by a constant amplitude $b_{\mbox{in}}(t)=b_{\mbox{in}}$, 
the stationary
solution of the atom-field dynamics reads
\begin{eqnarray}
\label{eq:stat}
\langle \hat{\sigma}_z \rangle &=& 
\frac{- \Gamma}{2 \Gamma +8 |b_{\mbox{in}}|^2} 
\nonumber \\
\langle \hat{\sigma}_- \rangle &=& 
\frac{- \sqrt{2\Gamma} \; b_{\mbox{in}}}{\Gamma +4 |b_{\mbox{in}}|^2}
\nonumber \\
b_{\mbox{out}} &=& \left(1-\frac{2\Gamma}{\Gamma +4 |b_{\mbox{in}}|^2}\right)
b_{\mbox{in}}.
\end{eqnarray}
The nonlinearity of the response function is given by the
ratio of $b_{\mbox{out}}$ and $b_{\mbox{in}}$,
\begin{equation}
\label{eq:nonlin}
\frac{b_{\mbox{out}}}{b_{\mbox{in}}}
= \frac{4 |b_{\mbox{in}}|^2/\Gamma - 1}{4 |b_{\mbox{in}}|^2/\Gamma + 1}.
\end{equation}
Figure \ref{response} illustrates this nonlinearity of the atomic 
response function. 
Experimentally, this response function can be directly observed
using homodyne detection. Note that this was also done in 
\cite{Tur95}, but the data was separated into a transmitted 
amplitude and a phase. In our case, this separation would 
create the impression of a sudden jump from a phase shift of
$\pi$ associated with a negative response function to a phase
shift of zero associated with a positive response function
at a switching intensity of $|b_{\mbox{in}}|^2 = \Gamma/4$. 
The resonant configuration shown in figure \ref{geometry} 
therefore allows an optimization of the phase shift observed 
in \cite{Tur95} to the ideal case of a seemingly instantaneous 
phase change of $\pi$. 

\begin{figure}
\begin{picture}(320,200)
\put(60,30){\makebox(250,150){\includegraphics{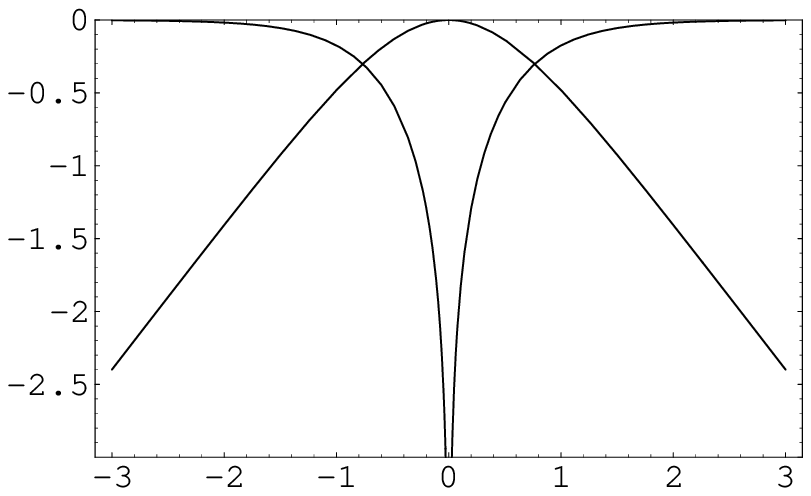}}}
\put(0,86){\makebox(60,20){\large
$\log_{10} \left[\frac{|b_{\mbox{\small out}}|^2}
{|b_{\mbox{\small in}}|^2}\right]$}}
\put(0,114){\makebox(60,20){\large
$\log_{10} \left[\frac{P_{\mbox{\small noise}}}
{|b_{\mbox{\small in}}|^2}\right];$}}

\put(155,65){\makebox(20,20){$\log_{10} 
\left[\frac{|b_{\mbox{\tiny out}}|^2}
{|b_{\mbox{\tiny in}}|^2}\right]$}}
\put(178,88){\vector(1,1){12}}
\put(115,145){\makebox(20,20){$\log_{10} 
\left[\frac{P_{\mbox{\tiny noise}}}
{|b_{\mbox{\tiny in}}|^2}\right]$}}
\put(128,142){\vector(1,-1){12}}

\put(165,10){\makebox(60,20){\large $\log_{10} 
\left[\; 4 |b_{\mbox{\small in}}|^2 / \Gamma \;\right] $}}
\end{picture}
\caption{\label{noise} Relative contributions of the coherent
signal intensity $|b_{\mbox{out}}|^2$ and the incoherent noise
$P_{\mbox{noise}}$ in the output as a function of scaled input 
intensity $4 |b_{\mbox{in}}|^2 / \Gamma$.}
\end{figure}

However, equation (\ref{eq:nonlin}) also indicates that the
coherent amplitude of the response changes during the switching
process. 
This change of amplitude is due to the 
incoherent emission $P_{\mbox{noise}}$ described by equation 
(\ref{eq:noise}). As shown by the noise limit (\ref{eq:limit}),
this incoherent contribution is unavoidable when the nonlinearity
originates from the saturation of a two level atom.
The relative contributions of
coherent response and incoherent noise to the total output
intensity are given by
\begin{eqnarray}
\left|\frac{b_{\mbox{\small out}}}{b_{\mbox{\small
in}}}\right|^2 &=&
\frac{(4 |b_{\mbox{in}}|^2/\Gamma - 1)^2}{(4 |b_{\mbox{in}}|^2/\Gamma + 1)^2}
\nonumber \\[0.3cm]
\frac{P_{\mbox{\small noise}}}{|b_{\mbox{\small in}}|^2} &=&
\frac{16 |b_{\mbox{in}}|^2/\Gamma}{(4 |b_{\mbox{in}}|^2/\Gamma + 1)^2}.
\end{eqnarray}
Figure \ref{noise} illustrates this intensity dependence of the
noise in the output field. The phase flip at $|b_{\mbox{in}}|^2 = \Gamma/4$
is clearly associated with a complete phase randomization
in the emitted field. A look at the steady state described by
equation(\ref{eq:stat}) shows that the coherent dipole is 
reduced to one half its linear value at this intensity. 
As a result, the amplitudes of dipole emission and of directly
reflected light are exactly equal and the coherent output 
component is eliminated by the destructive 
interference of these two amplitudes. The remaining emission
at $|b_{\mbox{in}}|^2 = \Gamma/4$ is therefore a completely
incoherent random phase emission originating entirely from 
quantum fluctuations in the light-atom interaction.
The gradual change in amplitude shown in figure \ref{response}
is thus associated with a rather drastic increase of noise
at the switching point. Note that this kind of noise increase
at the switching threshold is also typical for classical
bistable systems. The quantum mechanical properties of a single 
two level system thus appear to be very similar to the noise 
properties of classical nonlinear systems. 

A fully coherent output is only obtained in the
linear limit given by 
$|b_{\mbox{in}}|^2 \ll \Gamma/4$, and in the fully saturated 
limit, given by $|b_{\mbox{in}}|^2 \gg \Gamma/4$.
It is therefore necessary to consider the sensitivity of
various applications to this noise effect. In particular, 
there is a great difference between an application to 
interference effects with coherent light
and the single photon switching envisioned in
\cite{Tur95,Nie}, since the interference measurments allow
an averaging of the signal, whereas single photon experiments
do not allow this. 
The adiabatic elimination used in our model suggests that
the average photon number in the cavity is not a useful
criterion of the strength of the nonlinearity at the quantum
level. Rather, it is more appropriate to compare the
rate of incoming photons with the spontaneous emission rate
of the excited atom. 
In the configuration discussed here, the switching
intensity $|b_{\mbox{in}}|^2 = \Gamma/4$ is only one eighth of the 
spontaneous emission rate. This means that on average, only
one eighth of a photon arrives at the atom during the spontaneous
emission lifetime $\tau_{\mbox{sp.}}=1/(2\Gamma)$. 
Since switching is possible at such a low photon density,
it may be possible to use the dissipation free two level
nonlinearity to implement a conditional phase shift of $\pi$
for exactly two photons. Specifically, the average intensity
$2/T$ of a two photon pulse of pulse duration $T$ would 
exceed the switching intensity for pulse durations of 
$T<8/\Gamma$. This pulse length may indeed be sufficient to 
justify the analogy between the coherent input field and the
two photon state claimed in \cite{Tur95}. However, it should 
be noted that the switching noise in the two photon
interaction will reduce the single photon coherence.
A detailed investigation of this effect therefore requires 
a fully quantum mechanical treatment of the spatiotemporal
two photon wavefunction \cite{Koj02,Hof03}.

In general, an experimental realization of the 
dissipation free nonlinearity described in this
paper requires that losses to non-cavity modes
are negligible ($\gamma \ll 2 \Gamma$).
In order to understand how restrictive this limitation
is, it may be useful to consider the modification 
of the linear response of the atom in the presence of
losses. These losses can be included in the Bloch 
equations (\ref{eq:bloch}) by increasing the decay rates 
of the dipole and the inversion to $\Gamma+\gamma/2$ and 
to $2\Gamma+\gamma$, respectively.
This increase can also be represented by the spontaneous 
emission ratio $\beta = \Gamma/(\Gamma+\gamma/2)$, 
which is defined as the fraction of spontaneous emission 
that is emitted through the cavity mode. 
The total spontaneous emission rate of the atom is then
given by $ 2\Gamma/\beta$. In the linear case, the
effect of this increased damping rate in the atomic
response simply reduces the dipole response by a 
factor of $\beta$. According to equation (\ref{eq:inout}),
the linear output field amplitude is then given by  
\begin{equation}
\label{eq:beta}
b_{\mbox{out}} = (1-2 \beta) b_{\mbox{in}}.
\end{equation}
Thus, spontaneous emission into non-cavity modes gradually 
convert the phase shift of $\pi$ caused by the resonant 
dipole response back into absorption losses \cite{note}. 
The linear phase shift of $\pi$ can be observed as
long as more than half of the spontaneous emission is 
emitted through the cavity. Since the saturation of the
two level atom will eventually result in a phase shift 
of zero, a nonlinear phase change from phase $\pi$ to phase zero 
can be obtained whenever the linear phase shift 
is equal to $\pi$. 
The condition for an experimental observation of the 
nonlinear phase change is therefore
the realization of optical confinement with $\beta>1/2$.
At present, the most promising 
technology for the experimental realization of the dissipation 
free nonlinear phase shift seems to be the microcavity system 
presented in \cite{Tho98,Tur95}. In particular, the parameters 
given in \cite{Tur95} correspond to a value of 
$\beta \approx 0.7$. If the setup illustrated in figure 
\ref{geometry} were to be realized using exactly the same 
elements as in \cite{Tur95}, this should already be sufficient 
to observe the phase flip described in this paper. However, 
$\beta \approx 0.7$ still implies that more than $80\%$ of
the input photons are lost in the linear response case, so
further improvements of cavity geometries may be needed
for an implementation of a nearly dissipation free device.
Moreover, the reflection geometry proposed in this paper
requires that special care should be taken to achieve 
proper mode matching between the cavity mode and the pump 
beam. In principle, this problem is very similar to the
mode matching required to match the local oscillator beam 
to the cavity output in the transmission geometry. 
However, the effect of additional reflections of mismatched
pump light may increase the effect of mode matching errors.
The reflection geometry proposed here could therefore be 
more challenging to realize than the transmission geometry
reported in \cite{Tho98,Tur95}. Nevertheless our theoretical 
predictions for this geometry indicate that the realization
of such a geometry may well be worth the effort.

In conclusion, we have shown that the maximal nonlinear
phase shift of $\pi$ can be obtained by using a single 
two level atom in a one sided cavity with negligible losses. 
It is then possible to use resonant light to obtain a 
nonlinearity sensitive to individual excitations of the
atom. The reflection geometry illustrated in figure 
\ref{geometry} thus optimizes the nonlinear phase shifts 
previously obtained in the experiment
by Turchette et al. \cite{Tur95}.

Part of this work was supported by the program "Research and
Development on Quantum Communication Technology" of the
Ministry of Public Management, Home Affairs, Posts and
Telecommunications of Japan.



\begin{thebibliography}{xyz00}
\bibitem{Foot}
Note that the reflected light initially overlaps completely 
with the input beam. However, the separation of two 
counterpropagating beams can be achieved by using standard
optical elements such as optical directional couplers.
\bibitem{note}
It may be worth noting that absorption always corresponds to
destructive interference between the phase flipped dipole 
emission stimulated by the input field and the input field itself.
In this sense, the phase shift properties of resonant absorption
are quite universal.
\end{thebibliography}
\end{document}